\documentclass[preprint]{aastex}
\usepackage{mkfig}

\begin{document}
 
\title{\bf Measured Mass Loss Rates of Solar-like Stars as a Function of
  Age and Activity\altaffilmark{1}}

\author{Brian E. Wood\altaffilmark{2}, Hans-Reinhard
  M\"{u}ller\altaffilmark{3,4}, Gary P. Zank\altaffilmark{4}, Jeffrey L.
  Linsky\altaffilmark{2}}

\altaffiltext{1}{Based on observations with the NASA/ESA Hubble Space
  Telescope, obtained at the Space Telescope Science Institute, which is
  operated by the Association of Universities for Research in Astronomy,
  Inc., under NASA contract NAS5-26555.}
\altaffiltext{2}{JILA, University of Colorado and NIST, Boulder, CO
  80309-0440; woodb@origins.colorado.edu, jlinsky@jila.colorado.edu.}
\altaffiltext{3}{Bartol Research Institute, University of Delaware, Newark,
  DE 19716; mueller@bartol.udel.edu.}
\altaffiltext{4}{Institute of Geophysics and Planetary Physics,
   University of California at Riverside, 1432 Geology, Riverside, CA 92521;
   zank@ucrac1.ucr.edu.}

\begin{abstract}

     Collisions between the winds of solar-like stars and the local ISM
result in a population of hot hydrogen gas surrounding these stars.
Absorption from this hot H~I can be detected in high resolution Ly$\alpha$
spectra of these stars from the {\em Hubble Space Telescope}.  The amount of
absorption can be used as a diagnostic for the stellar mass loss rate.
We present new mass loss rate measurements derived in this fashion for four
stars ($\epsilon$~Eri, 61~Cyg~A, 36~Oph~AB, and 40~Eri~A).  Combining
these measurements with others, we study how mass loss varies with stellar
activity.  We find that for the solar-like GK dwarfs, the mass loss per
unit surface area is correlated with X-ray surface flux.  Fitting a power
law to this relation yields $\dot{M}\propto F_{x}^{1.15\pm 0.20}$.  The
active M dwarf Proxima~Cen and the very active RS~CVn system $\lambda$~And
appear to be inconsistent with this relation.  Since activity is known to
decrease with age, the above power law relation for solar-like stars
suggests that mass loss decreases with time.  We infer a power law
relation of $\dot{M}\propto t^{-2.00\pm 0.52}$.  This suggests that
the solar wind may have been as much as 1000 times more massive in the
distant past, which may have had important ramifications for the history
of planetary atmospheres in our solar system, that of Mars in particular.

\end{abstract}


\section{INTRODUCTION}

     The existence of a mass outflow from the Sun has been recognized for
many decades, its presence revealed most visibly by the behavior of comet
tails \citep[e.g.,][]{lb57}.  It has also been recognized for many years
that the solar wind has its origin in the Sun's hot corona.  Thus, it is
expected that all cool stars with analogous hot coronae will have similar
winds \citep[e.g.,][]{enp60}.

     Nevertheless, actually detecting a solar-like wind emanating from
another star has proved to be very difficult.  This is perhaps not
surprising given that the solar wind itself is not particularly easy to
observe remotely.  The low density of the wind, corresponding to a mass
loss rate of only $\dot{M}_{\odot}=2\times 10^{-14}$ M$_{\odot}$ yr$^{-1}$
\citep[e.g.,][]{wcf77}, and its high temperature and ionization state
make it difficult to detect with simple imaging or spectroscopic techniques.
Trying to detect such a wind around another star is needless to say even
more daunting.  This is in contrast to the massive winds of hot stars and
evolved cool stars, which are comparatively easy to detect
spectroscopically thanks to the characteristic P-Cygni line profiles that
signify the presence of these strong winds.

     Fortunately, a method for indirectly detecting solar-like stellar
winds has recently become available using high resolution {\em Hubble Space
Telescope} (HST) spectra of the H~I Ly$\alpha$ lines from nearby stars.  The
method is indirect because the material that is detected in Ly$\alpha$
absorption is not from the wind itself, which is fully ionized and therefore
has no H~I, but is instead interstellar H~I that is heated within the
interaction region between the wind and the local interstellar medium
(LISM).  The existence of this
detectable hot H~I was first suggested by hydrodynamic models of our own
heliosphere, which describe how the radial solar wind interacts with the
laminar interstellar wind seen from the Sun's rest frame
\citep{vbb93,vbb95,hlp95,gpz96,vbb98,gpz99}.

     The plasma collisions between the fully ionized solar wind and
partially ionized LISM establish the familiar large scale structure of the
heliosphere, consisting of three discontinuities in the flow:  the
termination shock where the solar wind is shocked to subsonic speeds,
the heliopause separating the plasma flows of the wind and the LISM, and the
bow shock where the interstellar wind is shocked to subsonic speeds.  The
distances to these boundaries in the upwind direction are currently
estimated to be roughly 85, 125, and 240 AU for the termination shock,
heliopause, and bow shock, respectively.  The {\em Voyager}~1 and
{\em Voyager}~2 spacecraft are currently 82 and 65 AU away, respectively,
more or less in the upwind direction, so hopefully they will soon cross the
termination shock and observationally establish its true location.

     The LISM is only partially ionized, but including the neutral
LISM component in models of the heliosphere is difficult, because the charge
exchange processes that allow the H~I to take part in the wind-LISM
interaction force the H~I out of thermal and ionization equilibrium.  It is
only relatively recently that codes including neutrals in a self-consistent
manner have been developed \citep*{vbb93,vbb95,hlp95,gpz96}.
It is these models that show that our
heliosphere should be permeated by H~I with an effective temperature of
$T\sim 20,000-40,000$~K, much of which is piled up between the heliopause
and bow shock, creating a so-called ``hydrogen wall.''  It is this hot H~I
that produces a Ly$\alpha$ absorption signature broad enough to be at least
partially separable from the absorption of the cooler LISM material
($T\approx 8000$~K).

     This hot heliospheric H~I was first detected in Ly$\alpha$
observations of $\alpha$~Cen A and B \citep{jll96,kgg97},
where it was shown that the ISM alone could not account for the
observed H~I absorption, and that the excess absorption on the red side of
the line was consistent with the heliospheric H~I absorption predicted by
the heliospheric models.  The heliospheric absorption is redshifted
relative to the ISM due to the deceleration of ISM material as it crosses
the bow shock.  The heliospheric models could {\em not}, however, account
for the observed excess on the blue side of the line, which \citet{kgg97}
attributed to analogous ``astrospheric'' H~I surrounding
$\alpha$~Cen~A and B.  This interpretation was confirmed by observations of
$\alpha$~Cen's distant companion Proxima~Cen, which show that the blue side
excess is {\em not} seen toward Proxima~Cen, proving that the blue side
excess absorption seen toward $\alpha$~Cen must be due to circumstellar
material around $\alpha$~Cen that does not extend as far as the distant
companion Proxima~Cen \citep{bew01}.  In the meantime, astrospheric
absorption was also detected toward several other nearby cool stars:
$\epsilon$~Ind and $\lambda$~And \citep{bew96};
$\epsilon$~Eri \citep{ard97}; 61~Cyg~A and 40~Eri~A \citep{bew98};
and 36 Oph \citep{bew00a}.  However, the 40~Eri~A
and $\lambda$~And detections must be regarded as tentative and require
confirmation.

     These detections of astrospheric H~I represent the first detections,
albeit indirect, of solar-like winds around other stars.  Detecting
solar-like stellar winds is important for many reasons.  By observing solar
phenomena such as the solar wind on other stars, we can in principle
improve our understanding of these phenomena in ways that are impossible
based on observations of the Sun alone.  For example, the extensively
studied correlation between rotation and stellar magnetic activity reveals
the importance of rapid rotation in the generation of high activity levels
in stars \citep{rp81,fmw82,fmw83,jpc85,gm85,taf89,jrs94,tra97}.
The rotation/activity correlation provides
valuable information about the dynamo mechanism that produces magnetic
activity on the surfaces of cool stars, including the Sun, but this
information can only be acquired by observing stars with different rotation
rates.  Similarly, our understanding of the acceleration mechanisms of
solar-like winds could in principle be improved significantly by
measurements of how wind properties differ for different types of cool
stars.  Since winds play a crucial role in the magnetic braking process
that slows stellar rotation with time, measuring wind properties of stars
with different ages could also improve our understanding of stellar angular
momentum evolution.

     However, in order for these benefits to be realized, the astrospheric
analyses must not only yield detections of stellar winds but also
measurements of their properties.
Fortunately, the amount of astrospheric absorption is dependent on the
stellar mass loss rate, since larger mass loss rates will yield larger
astrospheres and therefore larger column densities.  With the assistance
of hydrodynamic models, \citet{bew01} demonstrated that the
astrospheric absorption observed toward $\alpha$~Cen suggests a mass loss
rate of $\dot{M} = 2~\dot{M}_{\odot}$ for the combined winds of
$\alpha$~Cen~A and B (both stars lie within the same astrosphere), and an
upper limit of $\dot{M} < 0.2~\dot{M}_{\odot}$ for Proxima~Cen.
\citet{hrm01a} presented estimates for the mass loss rates of
$\epsilon$~Ind and $\lambda$~And, which were revised by \citet{hrm01b}
based on additional modeling to $\dot{M} = 0.5~\dot{M}_{\odot}$ for
$\epsilon$~Ind and $\dot{M} = 5~\dot{M}_{\odot}$ for $\lambda$~And.  In
this paper, we estimate mass loss rates for the other stars with detected
astrospheric absorption ($\epsilon$~Eri, 61~Cyg~A, 40~Eri~A, and 36 Oph).
Preliminary results of this analysis were presented by \citet{bew02}.
Finally, we consider all the mass loss measurements together to see what
they suggest about how mass loss varies with activity and spectral type.

\section{MEASURING MASS LOSS RATES}

\subsection{The Sample of Stars with Detected Astrospheres}

     The seven stars with detected astrospheric absorption are listed in
Table~1, along with their spectral types and {\em Hipparcos} distances
\citep{macp97}.  Also listed is Proxima~Cen, for which an upper
limit for astrospheric absorption yields an upper limit for its mass loss
rate \citep{bew01}.  In some binary systems, both stars will lie
within the same astrosphere, meaning the astrospheric absorption will be
characteristic of the combined mass loss of both stars.  This is the case
for the $\alpha$~Cen, 36~Oph, and $\lambda$~And systems (see Table~1),
while 61~Cyg~A and 40~Eri~A will have astrospheres all to themselves
despite being members of multiple star systems (see \S2.4--2.6).

     The surface areas listed in Table~1 (in solar units) are the combined
surface areas of all stars within the detected astrospheres.  Most of these
areas are based on stellar radii computed using the Barnes-Evans relation
\citep{tgb78}, except for Proxima Cen and $\lambda$~And,
for which we assume radii of 0.16~R$_{\odot}$ and 7.4~R$_{\odot}$,
respectively \citep{pmp93,ten99}.  The
activity level of the stars is indicated in Table~1 by their X-ray
luminosities, $L_{x}$, in units of ergs~s$^{-1}$.  These luminosities are
based on ROSAT PSPC data.  Most of the $\log L_{x}$ values are from
\citet{mh99}, with the exception of the $\lambda$~And
luminosity, which is from \citet{ao97}.  Following \citet{js95}, we assume
that 61~Cyg~A contributes 64\% of the binary's X-ray flux.

     In order to use the observed amount of astrospheric absorption to
derive a mass loss rate, it is first necessary to estimate the interstellar
wind velocity seen by each star ($V_{ISM}$) and the orientation of the
astrosphere relative to the line of sight, which can be described by the
angle between the upwind direction and the line of sight ($\theta$).  Two
velocity vectors must be known in order to compute these quantities, that
of the interstellar flow and that of the star itself.  Since all the
stars in Table~1 are very nearby, their proper motions, radial velocities,
and thus their stellar vectors are known very well.

     For most of the stars, we simply use the velocity vector of the Local
Interstellar Cloud (LIC) from \citet{rl95} to represent the ISM
flow at the location of the star.  Two exceptions are the Alpha/Proxima Cen
and 36~Oph systems, which lie not within the LIC but within the ``G cloud,''
which has a vector measured by \citet{rl92}.  The G cloud
vector is, however, very similar to the LIC vector, so its use results only
in minor modifications to the $V_{ISM}$ and $\theta$ values.  A few of the
stars in Table~1 lie in neither the LIC or G clouds.  The 61~Cyg system
clearly falls in this category based on the two-component velocity
structure seen for this line of sight \citep{bew98}, and the
$\lambda$~And system is much too far away to be within the LIC or G clouds.
However, even when multiple ISM components are detected within the
LISM, they are generally separated by no more than $5-10$ km~s$^{-1}$,
meaning the LIC vector should be a reasonable approximation for these other
nearby clouds, much as it is for the G cloud.  In any case, the $V_{ISM}$
and $\theta$ values for all stars in our sample are listed in Table~1.  The
$V_{ISM}$ values can be compared with the solar value of 26 km~s$^{-1}$
suggested by the LIC vector.

     Note that two of the astrospheric detections have been flagged as
being uncertain in Table~1.  The $\lambda$~And detection is uncertain
because there are no observations of narrow ISM lines such as Mg~II h \& k
to provide information on the velocity structure of the ISM for that line
of sight.  As a consequence, the H~I and D~I (deuterium) Ly$\alpha$ lines
are analyzed assuming a single absorption component.  This analysis clearly
suggests the presence of excess H~I absorption on the blue side of the
line that \citet*{bew96} interpret as astrospheric absorption, but
without knowledge of the ISM velocity structure it remains possible that
there is a blueshifted ISM component that might be able to account for the
H~I excess without requiring an astrospheric component.  The
40~Eri~A detection is also flagged as uncertain for reasons that will be
discussed in \S2.6.

     The third-to-last column of Table~1 lists the column density of the
hot astrospheric H~I based on the best astrospheric model fit to the data,
and the next column is the mass loss rate assumed for this best model.
Finally, the last column lists references relating to the detection of the
astrospheric absorption and from it the measurement of the mass loss rate.
However, as noted in \S 1, mass loss rates for four of the stars
($\epsilon$~Eri, 61~Cyg~A, 40~Eri~A, and 36 Oph) are measured for the first
time in this paper.  Thus, we now describe in detail how we make these
measurements.

\subsection{Astrospheric Modeling}

     One of the best examples of astrospheric absorption is that observed
toward $\epsilon$~Eri.  Figure~1 shows the Ly$\alpha$ profile of
$\epsilon$~Eri observed by the Goddard High Resolution Spectrograph (GHRS)
instrument on HST.  Like all such profiles, the center of the emission line
is absorbed by a very broad H~I absorption feature flanked by a much
narrower absorption feature from interstellar D~I, $-0.33$~\AA\ from
the H~I absorption.

     The upper solid line in Figure~1 is an estimate of the original
stellar Ly$\alpha$ emission profile.  Uncertainty in the stellar
Ly$\alpha$ profile is the dominant source of systematic error in the
analysis of the H~I absorption for all the stars in Table~1.  Thus, a
major goal of the empirical analyses referenced in Table~1 is to
experiment with different profiles to make sure that results are not
dependent on the assumed profile.  The Mg~II h \& k line profiles
(at 2803.531~\AA\ and 2796.352~\AA, respectively) are sometimes used
as guides for how the stellar profiles should appear, since the Ly$\alpha$
and Mg~II lines are both highly optically thick chromospheric lines and
the solar Ly$\alpha$ and Mg~II profiles are known to be quite similar.
This is why most of the profiles for the stars in Table~1 are assumed to
have self-reversals near line center, like that for $\epsilon$~Eri in
Figure~1, although results of the H~I absorption fitting are never very
sensitive to the exact depth or shape of the self-reversal.  Some analyses
also force the wings of the Ly$\alpha$ line to be centered on the rest
frame of the star, which can be a useful constraint in certain
circumstances \citep*{bew98,bew00a}.  We refer
the reader to the references listed in Table~1 for details about each
particular analysis, but considering possible alterations to assumed
stellar profiles is also necessary here in deciding if an astrospheric
model is consistent with the data or not (see below and \S2.6).

     The dotted line in Figure~1 is the profile after ISM
absorption alone, which is derived by fitting the
Ly$\alpha$ absorption profile while using the D~I absorption to constrain
the properties of the H~I absorption.  The centroid velocity of the H~I
absorption is forced to be the same as that of the D~I absorption, and since
H~I and D~I lines in the LISM are thermally broadened, the Doppler
broadening parameters of the H~I and D~I absorption can also be related by
$b(H~I)=\sqrt{2} b(D~I)$ \citep*[see, e.g.,][]{jll96,bew98,bew00a}.
With these sensible constraints, it becomes
immediately obvious that the H~I absorption is inconsistent with the D~I
absorption.  In particular, it is too broad and blueshifted relative to D~I.
Thus, there is a large excess of absorption on the blue side of the H~I
absorption that cannot be explained by ISM absorption \citep{ard97}.

     Hydrodynamic models of the heliosphere suggest that heliospheric
H~I absorption should be redshifted relative to the ISM absorption
regardless of whether the line of sight is upwind, downwind, or sidewind.
This is primarily due to the deceleration and deflection of interstellar
material as it crosses the bow shock, although other factors are at work
downwind \citep{vvi99,bew00b}.
Conversely, astrospheric absorption should generally be
{\em blueshifted} relative to the ISM absorption, since we are observing
the absorption from outside the astrosphere rather than inside.  All of the
excess absorption observed toward $\epsilon$~Eri is on the blue side of the
line, suggesting that the stellar astrosphere is responsible for the excess
rather than our heliosphere.

     To quantitatively measure mass loss rates based on astrospheric
absorption like that seen for $\epsilon$~Eri in Figure~1 requires the
use of hydrodynamic models of the astrosphere.  However, it is
crucial that these astrospheric models be extrapolated from a
heliospheric model that has been demonstrated to successfully reproduce
heliospheric H~I absorption.  Essentially, the heliospheric H~I absorption
is used to calibrate the models before applying them to the case of
astrospheres.

     Excess H~I absorption on the red side of the Ly$\alpha$ line that can
be interpreted as heliospheric in origin has been clearly detected toward
three different stars:  $\alpha$~Cen \citep{jll96}, 36~Oph \citep*{bew00a},
and Sirius \citep*{vvi99}.  There are more subtle
suggestions of some heliospheric contribution to the H~I absorption
observed toward Capella and G191-B2B, which are very similar lines of
sight \citep{avm98,ml02}.  \citet*{bew00b}
compare the heliospheric absorption observed for the first three
lines of sight listed above (and upper limits for heliospheric absorption
provided by three other lines of sight) with the predictions of many
heliospheric models.  Two different codes are used, a
``Boltzmann'' code that uses a kinetic treatment of the neutrals \citep{hrm00},
and a ``four-fluid'' code that models the heliosphere
as an interaction involving four distinct fluids:  a plasma component and
three neutral components \citep{gpz96}.  The three neutral components
represent the three distinct regions within the heliosphere where charge
exchange occurs, which are separated by the heliopause and termination shock
boundaries.

     The solar wind parameters that must be assumed for these models are the
wind velocity ($V_{w}$), proton density ($n_{w}(H^{+})$), and temperature
($T_{w}$) at 1~AU; which are assumed to be $V_{w}=400$ km~s$^{-1}$,
$n_{w}(H^{+})=5$ cm$^{-3}$, and $T_{w}=10^{5}$~K.  These values are within
the observed range of solar wind properties at Earth  \citep[e.g.,][]{wcf77}.
Note that the actual solar wind varies with time and ecliptic
latitude, with $n_{w}(H^{+})=3$ cm$^{-3}$ and $V_{w}=750$ km~s$^{-1}$ being
more typical at high latitudes \citep{djm01}.  The
interstellar parameters that must be assumed include the ISM wind velocity
($V_{ISM}$), proton density ($n_{ISM}(H^{+})$), H~I density ($n_{ISM}(H~I)$),
and temperature ($T_{ISM}$).  The velocity $V_{ISM}=26$ km~s$^{-1}$ is known
very well for the Sun, but the other ISM parameters can be varied within
their various observational tolerances \citep*[see][]{bew00b}.
\citet*{bew00b} also experiment with various values for a parameter
increasing the ISM pressure in order to account for the highly uncertain
magnetic and cosmic ray pressure within the ISM.  Despite experimentation
with different input parameters, no Boltzmann model was found that could
adequately reproduce the heliospheric H~I absorption for all the lines of
sight mentioned above. However, a four-fluid model that assumes
$n_{ISM}(H^{+})=0.1$ cm$^{-3}$, $n_{ISM}(H~I)=0.14$ cm$^{-3}$,
$T_{ISM}=8000$~K, and no correction for ISM magnetic field or cosmic ray
pressure accurately reproduces the heliospheric absorption.

     Thus, we use this model as the starting point for our astrospheric
modeling.  This model is a $\dot{M} = 1~\dot{M}_{\odot}$ model appropriate
for a $V_{ISM}=26$ km~s$^{-1}$ system.  In order to construct a
$\dot{M} = 2~\dot{M}_{\odot}$ model for $\epsilon$~Ind, for example, we
would compute another four-fluid model with $n_{w}(H^{+})$ increased by a
factor of 2 to $n_{w}(H^{+})=10$ cm$^{-3}$ and with the ISM wind speed
changed to the $V_{ISM}=68$ km~s$^{-1}$ value appropriate for
$\epsilon$~Ind (see Table~1).  All other parameters are kept the same.

     This assumes that ISM parameters do not vary greatly from location to
location within the LISM.  This assumption is a source of systematic
error for our analysis, and we note that \citet{pcf93} actually proposed
using astrospheric properties as probes of the ISM rather than using
astrospheres as diagnostics for the stellar winds.  However, if different
regions of the LISM are in pressure equilibrium, LISM properties
should not vary too severely.  Furthermore, \citet{vvi02}
find that their models suggest surprisingly low sensitivity of
astrospheric H~I absorption to changes in the assumed ISM densities, at
least in upwind and sidewind directions, meaning that modest variations in
LISM densities should not greatly prejudice our results.

     The inaccuracies and different approximations in current
heliospheric/astrospheric models are such that different types of models
developed by different groups can be expected to predict somewhat different
properties and different amounts of H~I absorption, even if the models all
assume the same input parameters \citep*{vbb98,bew00b}.
This is why it is so important to extrapolate the astrospheric models from
a heliospheric model that successfully reproduces the observed heliospheric
H~I absorption.  If, for example, we were to find a heliospheric Boltzmann
model that reproduced the observed heliospheric H~I absorption, it would
likely have different input ISM parameters than those that worked for the
four-fluid code.  However, this does {\em not} necessarily mean that
stellar mass loss rates derived from Boltzmann models will be significantly
different from those derived here using the four-fluid code, as long as
the astrospheric Boltzmann models are extrapolated from the successful
heliospheric model, assuming the ISM input parameters that worked for
{\em that} model rather than the parameters that worked for the successful
{\em four-fluid} heliospheric model.  All types of models that currently
exist should have the property that larger mass loss rates will yield
increasing amounts of H~I absorption.  Thus, if all types of codes are
first calibrated by demonstrating that a $\dot{M} = 1~\dot{M}_{\odot}$,
$V_{ISM}=26$ km~s$^{-1}$ model reproduces the observed heliospheric H~I
absorption, much of the potential model dependence can be removed.

     Figure~2 shows examples of various astrospheric models constructed
for various stars in our sample.  These are in fact the models that yield
the best fits to the data, as we will show below.  The figure shows the
H~I density distribution.  In each panel, the high density area in red
between the bow shock and heliopause is the so-called
hydrogen wall that is responsible for most of the astrospheric absorption
for all of these lines of sight.  The models also provide distributions of
H~I temperature and flow velocity.  From these distributions, we can trace
out the H~I density, temperature, and flow velocity along the line of
sight to the star, which is indicated by the $\theta$ values in Table~1
and shown explicitly in Figure~2.  And from these tracings we can
determine the predicted H~I absorption based on the model.  We assume
that nonthermal velocities do not contribute significantly to the line
broadening.  This seems a safe assumption given that for the astrospheric
models in Figure~2, the high temperatures of the hot astrospheric H~I lead
to very large Doppler broadening parameters of $b=15-45$ km~s$^{-1}$,
meaning turbulence would have to be extreme to increase them even further.
Nonthermal broadening is known to be insignificant for H~I in the LISM,
where temperatures are lower \citep[e.g.,][]{bew98}.

     Technically, Figure~2 shows the H~I distribution for only the most
important of the three H~I components used in the four-fluid code, the
``Component 1'' neutrals \citep[see][]{gpz96}, which include the
undisturbed LISM H~I and H~I created by charge exchange with heated LISM
protons in the hydrogen wall.  The other two components are H~I particles
created by charge exchange with the very hot, decelerated solar wind
protons in between the termination shock and heliopause (``Component 2''),
and H~I created by charge exchange with the cold, fast solar wind protons
within the termination shock (``Component 3'').  We sum the absorption from
all three components before a final absorption profile is derived for the
model.

     By constructing different models with different assumed mass loss
rates, as described above, we can obtain predicted Ly$\alpha$ absorption
profiles for a range of mass loss rates to compare with the data.  This
comparison is made in Figure~3 for six of the stars in our sample.  Three
of the data-model comparisons in Figure~3 and best-fit models in Figure~2
($\epsilon$~Ind, $\alpha$~Cen, and $\lambda$~And) are actually from
previous work \citep*{bew01,hrm01b},
which we show here for purposes of comparison.  The other three stars
shown (61~Cyg~A, $\epsilon$~Eri, and 36~Oph) are new analyses.
We focus on the blue side of the H~I Ly$\alpha$ absorption feature in
Figure~3, since that is where the astrospheric absorption is most apparent.
Each panel shows the ISM absorption alone (green dashed line), derived from
previous empirical analyses (see references listed in Table~1) by forcing
the H~I absorption to be consistent with the D~I absorption, which is also
visible in the figure.  In all cases there is a substantial amount of
excess absorption that we interpret as astrospheric.  The H~I
absorption predictions of astrospheric models are shown as blue lines in
each panel.  Table~1 lists the mass loss rates that we believe lead to the
best matches to the data.

     In principle, discrepancies between the models and the data seen in
Figure~3 can be reduced if changes are made to the assumed stellar
Ly$\alpha$ profiles.  An example of this will be given in \S 2.6 for
40~Eri~A.  However, it is difficult to shift the location where the H~I
absorption becomes saturated by any {\em reasonable} change to the stellar
profile.  Trying to correct discrepancies at the base of the absorption
typically results in the introduction of unreasonable fine structure into
the assumed stellar line profile \citep*{bew00b}.  Thus, the base of
the H~I absorption (e.g., at $-50$ km~s$^{-1}$ for $\epsilon$~Ind or
$-25$ km~s$^{-1}$ for $\epsilon$~Eri in Fig.~3) is less affected by
uncertainties in the stellar profile than the upper part of the H~I
absorption profile, so the base of the profile is the best place to compare
the data and the models when deciding which model is the best fit.  For
this reason, we consider the best fit to the $\epsilon$~Ind data to be the
$0.5~\dot{M}_{\odot}$ model rather than the $0.8~\dot{M}_{\odot}$ model
(see Fig.~3).

     Figure~2 illustrates the large range of size scales that the
astrospheres have.  This is illustrated more directly in Figure~4,
where we show the variations of H~I density and temperature with distance
from the star in the upwind direction (i.e., $\theta=0^{\circ}$) for these
models and the best solar model from \citet*{bew00b}.  Stars with
faster $V_{ISM}$ speeds (e.g., $\epsilon$~Ind and 61~Cyg~A; see Table~1)
have hotter astrospheric H~I due to more heating at the bow shock.  Their
astrospheres also tend to be more compact, and the density enhancements
within the hydrogen wall somewhat larger.  Stars with larger mass loss
rates naturally tend to have larger astrospheres.

     We now discuss some aspects of the mass loss derivations for each
particular star separately, at least for the new analyses presented here.

\subsection{$\epsilon$ Eridani}

     The hydrodynamic models strongly suggest that blueshifted absorption
such as that seen toward $\epsilon$~Eri in Figure~1 is astrospheric
rather than heliospheric, but it is useful to rule out heliospheric
absorption empirically if possible.  As mentioned in \S 1, \citet{bew01}
demonstrated that the blue side excess absorption seen toward
$\alpha$~Cen could not be heliospheric (or interstellar) by comparing the
Ly$\alpha$ profiles of $\alpha$~Cen and Proxima~Cen and showing that the
blue side excess was not seen toward Proxima~Cen.  We can also demonstrate
empirically that the strong blue side excess absorption observed toward
$\epsilon$~Eri cannot be heliospheric absorption by comparing the
$\epsilon$~Eri data with Ly$\alpha$ observations of a similar line of sight
through the heliosphere.  The $\epsilon$~Eri line of sight is downwind
relative to the heliosphere, $\theta_{\odot}=148^{\circ}$ from the upwind
direction.  In Figure~5a, the $\epsilon$~Eri data are compared with the
HST/GHRS Ly$\alpha$ observations of 40~Eri~A, for which
$\theta_{\odot}=155^{\circ}$.  The 40~Eri~A data were analyzed by
\citet{bew98}; note that we have removed the geocoronal absorption from the
data in Figure~5a based on their analysis.  If the blue side excess
absorption seen toward $\epsilon$~Eri were heliospheric, there should be at
least that much absorption seen toward 40~Eri as well, but Figure~5a shows
that is clearly not the case, providing support for the astrospheric
interpretation of the excess absorption.

     Our technique for measuring mass loss rates from the astrospheric
absorption requires the assistance of hydrodynamic models of the
astrospheres, but for $\epsilon$~Eri it is actually possible to say
something about its mass loss rate relative to that of the Sun based on a
purely empirical comparison, thanks to the stellar $V_{ISM}$ value being
practically identical to the solar value of 26 km~s$^{-1}$ (see Table~1).
Furthermore, $\epsilon$~Eri is like the Sun in being within the LIC, since
the LIC absorption component is the only one detected toward $\epsilon$~Eri
\citep{ard97}.  Thus, if the $\epsilon$~Eri mass loss rate were
identical to that of the Sun, its astrosphere should have identical
properties to the heliosphere.

     We observe the astrosphere at an orientation angle of
$\theta=76^{\circ}$ (see Table~1).  In Figure~5b, the $\epsilon$~Eri data
are compared with HST/GHRS observations of 31~Com, for which
$\theta_{\odot}=73^{\circ}$.  There is no heliospheric absorption detected
toward 31~Com \citep*{ard97,bew00b}, but the amount of
absorption on the red side of the line provides an upper limit to the
amount of heliospheric absorption that can be present $73^{\circ}$ from the
upwind direction, which in Figure~5b is compared with the amount of
astrospheric absorption seen toward $\epsilon$~Eri at a similar
astrospheric angle.  In order to make the direct comparison in Figure~5b,
we have to reflect the $\epsilon$~Eri profile about the rest frame of the
interstellar medium and plot both spectra in that rest frame.  Figure~5b
clearly shows that $\epsilon$~Eri's astrosphere provides substantially more
absorption at $\theta=76^{\circ}$ than the heliosphere.  This suggests a
larger astrosphere containing more heated H~I, which can only
be the result of a larger mass loss rate, consistent with our measurement
of $\dot{M} = 30~\dot{M}_{\odot}$ using the astrospheric models (see
Fig.\ 3 and Table~1).

     Note the huge size of the $\epsilon$~Eri astrosphere suggested by the
best model in Figure~2, with an upwind bow shock distance of about 1600~AU.
The full width of the astrosphere is about 8000~AU based on its sidewind
extent.  At $\epsilon$~Eri's distance of 3.2~pc, this width corresponds to
an impressive angular size of $42^{\prime}$.  Thus, if $\epsilon$~Eri's
astrosphere were visible to the naked eye it would appear larger than the
full Moon!

     \citet{bew98} tried to estimate stellar wind ram pressures
from empirically measured astrospheric H~I column densities, using
arguments involving simple scaling laws.  For $\epsilon$~Eri, the wind was
actually estimated to be {\em weaker} than the Sun rather than stronger.
We believe that the analysis presented in this paper is greatly superior
to that attempted by \citet{bew98}, so we believe the dramatically
discrepant results for $\epsilon$~Eri indicate that the empirically
estimated astrospheric column density from \citet{ard97} is simply
not accurate enough to be able to infer stellar wind properties
from it.  The astrospheric H~I column density listed in Table~1 is in fact
over an order of magnitude higher than that estimated by \citet{ard97}.
The astrospheric absorption is a highly saturated absorption
feature in the flat part of the curve of growth, making its column density
very difficult to measure accurately even if the absorption was {\em not}
highly blended with the ISM absorption.  Furthermore, the scaling
arguments of \citet{bew98} do not take into account differences in
$\theta$ sampled by the lines of sight through the various astrospheres,
in contrast to the astrospheric modeling technique used here.  In the
final analysis, it is apparently necessary to use guidance from
astrospheric models in extracting wind properties from astrospheric
absorption.  The simpler technique attempted by \citet{bew98}
unfortunately does not seem to be reliable.

\subsection{61~Cygni~A}

     The K5~V star 61~Cyg~A is remarkable for having the most compact
astrosphere of those shown in Figures 2 and 4, with an upwind bow shock
distance of only about 30~AU.  This is comparable to the orbital distance
of Neptune in our solar system.  The compactness of the astrosphere is due
both to a very high ISM wind speed and a relatively low mass loss rate
(see Table~1).

     A K7~V companion star, 61~Cyg~B, exists only about
$30^{\prime\prime}$ away at a position angle of $150^{\circ}$ \citep{ah85}.
This corresponds to a projected distance of only 105~AU,
close enough that we were initally concerned that 61~Cyg~B might share an
astrosphere with 61~Cyg~A.  However, the size of the 61~Cyg~A
astrosphere in Figure~2 is small enough that the two stars should have
separate astrospheres.  We still must consider whether the tail of
61~Cyg~B's astrosphere happens to cross our line of sight to 61~Cyg~A.
Fortunately, we estimate that the downwind direction of the two 61~Cyg
astrospheres is at a position angle of $229^{\circ}$, roughly perpendicular
to a line between the two stars.  This means that the two astrospheres will
essentially sit side-by-side, and our line of sight to 61~Cyg~A should not
pass through the 61~Cyg~B astrosphere.  All this assumes that 61~Cyg~B's
astrosphere will not be larger than that of 61~Cyg~A, which seems a
reasonable assumption.

\subsection{36~Ophiuchi}

     The 36~Oph binary system consists of two K1~V stars in a highly
eccentric orbit with a semimajor axis of about $14^{\prime\prime}$,
corresponding to a separation of 77~AU \citep{awi96}.
Since this is much smaller than the size of the astrosphere shown in
Figure~2, both stars clearly lie within a common astrosphere.

     One unique aspect of the 36~Oph astrosphere is that it is the only
one detected for a downwind line of sight (i.e., $\theta>90^{\circ}$).
This makes the modeling harder because one must be careful to extend the
model grid far enough downwind to capture all the astrospheric H~I along
the observed line of sight.  Also, 36~Oph lies within the G cloud rather
than the LIC.  Since the G cloud is clearly cooler than the LIC based
on HST observations of $\alpha$~Cen and 36~Oph \citep*{jll96,bew00a},
when modeling the 36~Oph astrosphere we assume
the average G cloud temperature measured for these two analyses,
$T_{ISM}=5650$~K, rather than the LIC temperature of 8000~K.  We also use
the G cloud vector of \citet{rl92} to compute $V_{ISM}$,
rather than the LIC vector.  \citet{bew01} make similar assumptions
when modeling the $\alpha$~Cen astrosphere.

\subsection{40~Eridani~A}

     The K1~V star 40~Eri~A has two companion stars, both about
$80^{\prime\prime}$ away, corresponding to a separation of about 400~AU
\citep{kwk76}.  We do not show a 40~Eri~A astrospheric model in Figure~2,
but with an extremely high ISM wind speed of $V_{ISM}=127$ km~s$^{-1}$ and
a modest mass loss rate at best (see Table~1), the 40~Eri astrosphere
promises to be at least as compact as that of 61~Cyg~A, meaning that
40~Eri~A's companions will not be located within its astrosphere.

     In Figure~6a, we show the 40~Eri~A data and the assumed stellar
Ly$\alpha$ profile (solid line) from \citet{bew98}.  The dotted
line is the profile after ISM absorption is subtracted from the
stellar profile.  There is some evidence for a weak,
very broad excess of H~I absorption on the blue side of the line, which is
of a somewhat different character than the excesses seen in Figure~3.  The
extremely fast ISM wind speed seen by this star ($V_{ISM}=127$ km~s$^{-1}$)
would be expected to yield an extremely broad and unsaturated astrospheric
absorption profile, something like that suggested by Figure~6a, because
it should produce a very compact astrosphere with low H~I column densities
and very high temperatures.  However, the detection of the weak excess is
shaky \citep[see][]{bew98}, and is therefore flagged as
uncertain in Table~1.

     We also plot in Figure~6a the predicted astrospheric absorption for
models with $\dot{M} = 1~\dot{M}_{\odot}$ and $\dot{M} = 5~\dot{M}_{\odot}$
(dashed and dot-dashed lines, respectively).  Both models look like they
overpredict the amount of absorption, but one must be careful.
Unlike the examples in Figure~3, the astrospheric absorption is extremely
broad and unsaturated thanks to the large V$_{ISM}$ value.  This makes
it possible in principle to increase the fluxes of the assumed stellar
profile in order to greatly improve the agreement with the data, which is
done in Figure~6b.  As mentioned in \S2.2, the best place to compare
the models with the data is near the base of the absorption (at 1215.6~\AA),
as it is much harder to fix discrepancies at this location by changes to
the stellar profile.  The $1~\dot{M}_{\odot}$ fit is not as bad at that
location.  Figure~6b shows that it is possible to assume a stellar
Ly$\alpha$ profile that leads to much better agreement between the data
and the $1~\dot{M}_{\odot}$ model.  However, the discrepancies with the
data for the $5~\dot{M}_{\odot}$ are too severe for any reasonable changes
to the stellar profile to correct.  Thus, for 40~Eri~A we quote
$\dot{M}<5~\dot{M}_{\odot}$ (see Table~1), although since the
very detection of the astrospheric absorption is questionable, even this
upper limit must be regarded with some skepticism.

     The Ly$\alpha$ photons that we see absorbed by astrospheres are merely
scattered out of our line of sight rather than being destroyed.  Thus,
astrospheres are potentially sources of scattered Ly$\alpha$ {\em emission}
as well as absorption.  The compact size and high temperature expected for
40~Eri~A's astrosphere make it an attractive target to try to detect this
emission.  The compact size maximizes the surface brightness of
the astrosphere, and the high temperature should yield a broad emission
profile that should at least partially escape ISM absorption.  Thus, on
2001 January 26 an attempt was made to detect astrospheric Ly$\alpha$
emission surrounding 40~Eri~A, using the Space Telescope Imaging
Spectrograph (STIS) aboard HST.

     Geocoronal Ly$\alpha$ emission would drown
out the weak signal from the astrosphere in a simple UV image of the
system, so long slit spectroscopy was used to try to spectroscopically
separate the astrospheric Ly$\alpha$ emission from that of the geocorona.
We used a $52^{\prime\prime}\times 0.2^{\prime\prime}$ long slit and the
G140M grating to observe the Ly$\alpha$ spectral region, with an occulting
bar over the star to minimize scattered stellar light.  Unfortunately,
no extended astrospheric emission was detected in the 7814~s exposure.
Thus, these data fail to support the questionable detection of
astrospheric absorption from 40~Eri~A, although advanced radiative
transfer computations will be required to see if the nondetection of
emission is truly inconsistent with the amount of absorption suggested by
Figure~6a.  We save such computations for a future paper.

\section{SCIENTIFIC IMPLICATIONS OF THE MASS LOSS MEASUREMENTS}

\subsection{Mass Loss as a Function of Activity}

     The winds of cool main sequence stars have their origins in the
coronae of these stars, so it is natural to wonder whether mass loss rates
are correlated with coronal properties.  The coronal X-ray luminosity is a
good indicator of the magnetic activity level of a star and of the amount
of material that has been heated to high coronal temperatures.  Table~1
lists X-ray luminosities obtained with the {\em ROSAT} PSPC instrument,
assuming {\em Hipparcos} stellar distances.  Before comparing the
mass loss rates with the X-ray emission, it is necessary to make the
comparison equitable by dividing the mass loss measurements and X-ray
luminosities by the stellar surface areas (see Table~1).
Figure~7 shows mass loss rates per unit surface area plotted versus
X-ray surface flux.  For the solar-like GK dwarfs, the data suggest that
more active stars have higher mass loss rates.  However, the M dwarf
Proxima~Cen and the RS~CVn system $\lambda$~And (G8~IV-III+M~V) are
inconsistent with this mass-loss/activity relation.

     It has been argued previously that the strong flares on active M
dwarfs should induce large mass loss rates, which could have a large impact
on the Galactic ISM due to the large numbers of these stars \citep{djm99}.
But the Proxima~Cen data point in Figure~7 suggests that
active M dwarfs instead have {\em lower} mass loss rates than the
solar-like stars predict.  This may be telling us something important about
the wind acceleration process.  Why does it not work as well on the M
dwarfs?  Is it due to the higher surface gravity, the generally higher
coronal temperatures, stronger magnetic fields, or something else?
\citet{msg96} present evidence that M dwarfs have a somewhat
different magnetic field configuration than higher mass dwarfs, which could
also be a factor.  Addressing these issues will be an important challenge
for future models of the wind acceleration mechanisms for coronal stars.
However, there is only one M dwarf represented in Figure~7, and its mass
loss rate is only an upper limit, so additional observations of M dwarfs
are required to see whether Proxima~Cen's low mass loss rate is
representative of all active M dwarfs.

     The $\lambda$~And data point in Figure~7 is even more discrepant
from the solar-like stars than Proxima~Cen.  The $\lambda$~And system
is classified as an RS~CVn binary.  Such systems typically
have high rotation rates forced by tidal locking between
the two stars, which in turn leads to very high activity levels.  The
$\lambda$~And system certainly has the high activity, but is unusual in
having both a long rotation period of 54 days and a long orbital 
period of about 21 days \citep{lj96}, with the difference between the two
meaning that the two stars are {\em not} tidally locked.  The
$\lambda$~And primary, which surely dominates the X-ray flux and wind from
the system, has a surface area about 55 times that of the Sun.  The
estimated mass loss rate of $5~\dot{M}_{\odot}$ is therefore surprisingly
low, raising many of the same questions asked above regarding Proxima~Cen.
However, the uncertainty regarding the reality of the astrospheric
absorption precludes any detailed consideration of its mass loss behavior
at this time (see \S2.1).

     We fit a power law to the solar-like GK dwarfs in Figure~7, using a
Monte Carlo technique to estimate the best fit and its uncertainty.  In
order to estimate the uncertainty in the fit, we have to estimate
uncertainties for our mass loss rates.  Unfortunately, the mass loss
uncertainties depend entirely on systematic errors that are not easy to
quantify.  We now review some of these.

     We must assume in our modeling that the ambient LISM is the same
for our target stars, an assumption which is a
potential source of error.  However, as mentioned in \S2.2, the
requirements of pressure equilibrium suggest that LISM parameters should
not vary too severely among our stars.  Furthermore, Izmodenov et al.\
(2002) have found surprisingly little variability in heliospheric
H~I absorption in their models when different LISM parameters are assumed.
Uncertainties in V$_{ISM}$ and $\theta$ for each star will result in
uncertainties in the mass loss rate measurements (see \S2.2), but for these
nearby stars we believe the uncertainties in these quantities should not
be severe.

     We must assume that the astrospheric absorption that we see is
characteristic not only of the stellar mass loss along the line of sight,
but also characteristic for the star as a whole.  Solar wind properties
depend somewhat on ecliptic latitude, at least during solar minimum
conditions, so this is a potential source of error.  However, the solar
wind ram pressure variation with ecliptic latitude is not extreme (only
about a factor of 1.5), so three-dimensional models of the heliosphere that
take these variations into account do not suggest dramatic
latitude-dependent variations in heliospheric structure
\citep[e.g.,][]{hlp97}, so we do not think this is a big problem.

     Perhaps the greatest potential source of uncertainty is the assumption
that the stellar wind velocities are all close to $V_{w}=400$ km~s$^{-1}$.
We have no way of knowing how accurate this assumption might be, but we
note that the size of an astrosphere and the amount of astrospheric
absorption should scale roughly as the square root of the wind ram
pressure, $P_{w}$ \citep{bew98}.  Since $P_{w}\propto \dot{M} V_{w}$, our
mass loss estimates will to first order vary inversely with the assumed
wind speed.  Thus, if one believes that the wind velocity of a particular
star could be a factor of 2 different from 400 km~s$^{-1}$ (and solar wind
variations span a range of about a factor of 2),
our mass loss estimate for that star
could be off by a factor of 2.  Note that we could avoid this
issue to a large extent if we chose to measure and discuss wind ram
pressures rather than mass loss rates.  Regardless of what wind velocity we
choose to assume, the mass loss rate that we derive should be such that the
wind ram pressure is the same, since it is the ram pressure rather than the
wind velocity or mass loss rate alone that establishes the amount of
astrospheric absorption (as mentioned above).  However, the mass loss rate
is the quantity that is of most astrophysical interest, so that is what we
choose to discuss, even though this quantity cannot be measured as
precisely as $P_{w}$ due to the additional necessary $V_{w}=400$ km~s$^{-1}$
assumption.

     Finally, uncertainties in the intrinsic Ly$\alpha$ profiles lead to
uncertainties in the assessment of what model best fits the data (see
\S2.2 and Fig.~3).  Unlike the previously listed systematic errors, at
least in this case experimentation with the data provides us with some
idea of the magnitude of this uncertainty.  Considering all the possible
sources of error listed above, we estimate uncertainties for the mass
loss measurements
to be factors of 2 (i.e., $\pm 0.3$ dex).  These uncertainties are no
better than educated guesses, but we note that the assumed factor of 2
errors are roughly consistent with the scatter of the GK star data points
in Figure~7 about the best fit relation.  If the systematic errors are
really larger, one would expect to see more scatter, but a larger sample of
stars is necessary to address this issue.

       We also assume factor of 2 uncertainties for the X-ray luminosities,
not because of measurement errors but because of potential variability.
This assumption is consistent with the analysis of \citet{tra97}, who
estimated that the ROSAT PSPC fluxes for solar-like stars should cover a
mininum-maximum range of about a factor of 4 during activity cycles like
that of the Sun.  We note that X-ray luminosities measured from the
{\em Einstein} IPC instrument \citep{am87,js90} agree with the PSPC
luminosities in Table~1 to within a factor of 2 in all cases.

     The data points for the solar-like stars in Figure~7 are randomly
varied within the assumed error bars mentioned above for both the mass loss
rate and X-ray values, and a power law fit is performed for each trial.  The
line shown in Figure~7 is the average fit and the shaded region represents
the 1 $\sigma$ uncertainty in the fit.  In the figure, the relation is
extended up to the saturation line, which indicates the maximum X-ray flux
observed from solar-like stars \citep{mg97}.  The assumption is that when
coronal activity is at a maximum, coronal mass loss will be at a maximum as
well.  Quantitatively, the relation in Figure~7 is
\begin{equation}
\dot{M}\propto F_{x}^{1.15\pm 0.20}.
\end{equation}

\subsection{Mass Loss as a Function of Age}

     As stars age, their rotation rates ($V_{rot}$) slow down due to
magnetic braking.  As rotation slows, less magnetic activity is generated
by the stellar dynamo, so X-ray fluxes decrease.  There are a vast number
of observational studies that have been directed toward measuring the
relations between rotation and age \citep[e.g.,][]{as72,drs93}
and X-ray flux and rotation
\citep*[e.g.,][]{rp81,fmw82,fmw83,jpc85,gm85,taf89,jrs94}.
Quantitatively, \citet{tra97} estimates for
solar-like stars:
\begin{equation}
V_{rot}\propto t^{-0.6\pm 0.1}
\end{equation}
and
\begin{equation}
F_{x}\propto V_{rot}^{2.9\pm 0.3}.
\end{equation}
Combining equations (1)--(3), we can obtain the following relation for mass
loss and stellar age:
\begin{equation}
\dot{M}\propto t^{-2.00\pm 0.52}.
\end{equation}
This is the first empirically derived relation describing the mass loss
evolution of cool main sequence stars like the Sun.  The relation
suggests that mass loss decreases with time for solar-like stars.

     Note that the real relations between $V_{rot}$, $F_{x}$, and $t$
are probably not simple power laws.  Many of the references cited above
have assumed more complicated functional forms that might be more
physically realistic.  Nevertheless, we choose simple power law
representations from \citet{tra97} for the sake of convenience, since our
derived relation between $\dot{M}$ and $F_{x}$ in equation (1) is a simple
power law.  (Our small sample size does not warrant the assumption of a
functional form more complicated than a power law.)  If we were to choose
different relations from the literature, the quantitative form of
equation (4) would change.  However, all relations should yield the
same qualitative result that mass loss decreases with time.  For example,
if we assume
\begin{equation}
F_{x}\propto \exp \left[ \left( -2.20\pm 0.22 \right) t^{0.5} \right]
\end{equation}
from \citet{fmw91}, then we compute
\begin{equation}
\dot{M}\propto \exp \left[ \left( -2.53\pm 0.51 \right) t^{0.5} \right],
\end{equation}
where $t$ in equations (5) and (6) must be in Gyr units.  Equation (6)
appears very different from equation (4), but the differences
are not actually severe.  Equation (4) suggests that the solar mass
loss rate was $30-330$ times larger when the Sun was 1/10 its present
age, while equation (6) suggests that the solar wind was $20-90$ times
larger at that time.  There are differences between these two ranges, but
the disagreement is not excessive considering the uncertainties.

     The sample of stars that \citet{tra97} uses to derive relations
(2) and (3) has a range of age and activity similar to our sample in
Table~1, so combining equations (1)-(3) to obtain equation (4) is
appropriate.  It should be noted, however, that equation (2) clearly
does not work well for very young stars ($t<0.3$ Gyr).  Stars in young
clusters tend to show broad distributions of rotation rates, meaning that
the connection between age and rotation is not very strong for such stars
\citep{ts90,drs93}.  Thus, the predictions that equations (4) and (6)
make for the mass loss rates of very young stars should be considered
questionable.

     A theoretical point that should be made regarding equations (1)-(3)
is that while the connections between $V_{rot}$, $t$, and $F_{x}$ in
(2)-(3) are rather direct, the physical connection between $\dot{M}$ and
$F_{x}$ suggested by equation (1) must be indirect, because the X-ray flux
presumably originates in closed field regions while the stellar wind
arises in open field regions.  The apparent correlation between $\dot{M}$
and $F_{x}$ in Figure~7 may mean that the underlying cause of stronger
closed field regions that yield higher $F_{x}$ (i.e., the dynamo) also
yields stronger open fields that accelerate more wind.

     Figure~8 shows what relation (4) implies for the mass loss history of
the Sun, once again assuming that the mass loss ``saturates'' when the
coronal X-ray flux saturates.  The solar wind may have been $\sim 1000$
times stronger when the Sun was very young, although we note once again
that predictions for very early times are suspect.
The upper limits in Figure~8 are based on
nondetections of radio emission from three solar-like stars \citep{ejg00},
illustrating that the high mass loss rates predicted for young stars are
still consistent with our inability to detect radio emission from these
stars.  Note that the dependence of mass loss on time derived empirically
in relation (4) likely results from a number of physical processes,
including the evolution of magnetic fields in both strength and structure
(i.e., the relative importance of closed vs.\ open structure).

     Equation (4) has important ramifications for our understanding of
the angular momentum evolution of cool stars, because it is the interaction
between the stellar wind material and the magnetic field of the rotating
star that is believed to provide the braking mechanism for stellar rotation
on the main sequence, which is ultimately why rotation and activity both
decrease with time.  Higher mass loss rates can be expected to yield
stronger braking.  Models for how magnetic braking should affect stellar
rotation suggest relations of the form
\begin{equation}
\frac{\dot{\Omega}}{\Omega}\propto \frac{\dot{M}}{M}
  \left( \frac{R_{A}}{R} \right)^{m},
\end{equation}
where $\Omega$ is the angular rotation rate and $R_{A}$ is the Alfv\'{e}n
radius \citep*{ejw67,ks88,ejg00}.  The
exponent $m$ is a number between 0 and 2, where $m=2$ corresponds to a
purely radial magnetic field.  \citet{lm84} argues that more reasonable
magnetic geometries suggest $m=0-1$.

     We are interested in the time dependence of the quantities in equation
(7).  Equation (2) shows that $\Omega\propto V_{rot}\propto t^{-0.6}$, so
$\dot{\Omega}\propto t^{-1.6}$ and $\dot{\Omega}/\Omega \propto t^{-1}$.
We assume that neither the stellar mass ($M$) or radius ($R$) are time
variable (see \S3.3), while equation (4) indicates the time dependence of
$\dot{M}$.  The Alfv\'{e}n radius is
\begin{equation}
R_{A}=\sqrt{\frac{V_{w} \dot{M}}{B_{r}^{2}}},
\end{equation}
where $V_{w}$ is the stellar wind speed and $B_{r}$ is the radial magnetic
field.  Note that since equation (7) is computed assuming a simple, global
magnetic field geometry, $B_{r}$ should be considered a disk-averaged
field when relating it to the far more complex fields that stars like the
Sun actually have.  We assume that $V_{w}$ does not vary, which was also an
assumption used in the derivation of mass loss rates from the astrospheric
absorption (see \S2.2).  We can then express $B_{r}$ as a power law of
time, $B_{r}\propto t^{\alpha}$, where from equations (4), (7), and (8)
we find
\begin{equation}
\alpha = 1/m - \left( 1.00\pm 0.26 \right) \left( m+2 \right) /m.
\end{equation}
Assuming that $m$ is in the physically allowable range of $m=0-2$ yields
the upper limit $\alpha<-1.0$, while the more likely range of $m=0-1$
suggested by \citet{lm84} implies $\alpha<-1.2$.  In any case, this
result suggests that disk-averaged stellar magnetic fields decrease at
least inversely with age for solar-like stars.

\subsection{Relevance for Planetary Atmospheres}

     Figure~9 shows the cumulative mass loss for the Sun as a function of
time based on the relation in Figure~8.  Despite the high mass loss rates
predicted for the young Sun, the total mass that has been lost by the solar
wind is still $\lesssim 0.03$~M$_{\odot}$, not enough to have
dramatically altered the solar luminosity.  Thus, the higher mass loss
rates predicted for the young Sun are still not high enough to resolve the
so-called ``faint young Sun'' paradox, which arises from standard
evolutionary models of the Sun that predict that it should have been
about 25\% fainter 3.8 Gyr ago \citep{dog81}.  The faintness of the young
Sun appears to be inconsistent with the apparent existence of running
water at these times on the surfaces of both Earth and Mars, which implies
that planetary temperatures were not correspondingly lower
\citep{cs72,jfk91}.

     If the solar mass loss rate was high enough in the past to decrease
the Sun's mass by $\sim 10$\%, then the higher mass of the young Sun could
increase the predicted solar luminosity enough that it no longer is a
``faint young Sun'' \citep{jag87}.  Although our analysis does suggest higher
mass loss rates in the past, they are not high enough to change the solar
mass sufficiently to resolve the faint young Sun paradox, and Figure~9 shows
that almost all of the predicted mass loss occurs too early to be relevant to
the issue anyway.  A more likely resolution of the faint young Sun problem
lies in atmospheric greenhouse effects that could allow Earth and Mars to
maintain warm temperatures even with a fainter Sun \citep{jw85,jfk86,jfk91}.

     However, a more massive young solar wind may still have had profound
effects on planetary atmospheres in our solar system.  Exposure to the
solar wind can erode planetary atmospheres, and the higher mass loss
rates suggested for the young Sun in Figure~8 would exacerbate these
effects.  Solar wind sputtering processes have been proposed as having
important effects for the atmospheres of both Venus and Titan
\citep{ec97,hl00}, but the Martian atmosphere may be the most interesting
case of solar wind erosion, since the history of the Martian atmosphere is
linked with the issue of whether water, and perhaps life, once existed on
the surface.

     Unlike Earth, the Martian atmosphere is not currently protected from
the solar wind by a strong magnetosphere.  There is evidence that Mars once
had a magnetic field, which disappeared at least 3.9~Gyr ago \citep{mja99}.
At this point, the Martian atmosphere would have been exposed to a
solar wind about 40 times stronger than the current wind according to
Figure~8, which could have had a dramatic effect on the atmosphere.  Mars
appears to have had running water on its surface in the distant past, and
there is evidence from isotopic ratios in the Martian atmosphere that Mars
once had a thicker atmosphere that could have allowed a climate much more
conducive to the existence of surface water \citep[e.g.,][]{mhc96,bmj01}.
However, today the Martian atmosphere is very thin, making stable surface
water impossible, leading to the question of what happened to the thicker
atmosphere and surface water.  Solar wind erosion is a leading candidate
for the cause of this change \citep{jgl92,hp92,bmj94,dmk95,rl01}, and if
the young solar wind was stronger than today this possibility is even more
likely.

\section{SUMMARY}

     The interaction of a solar-like wind with the local ISM produces
a population of hot hydrogen gas that is detectable in high resolution
HST Ly$\alpha$ spectra of nearby stars.  The amount of astrospheric
H~I absorption is a diagnostic for the mass loss rate of the wind.
We have used mass loss rates measured from astrospheric absorption to
study the dependence of mass loss on stellar age and activity.  Our
results are as follows:
\begin{description}
\item[1.] We present new mass loss measurements for four stars with
  previously detected astrospheric absorption, using hydrodynamic models
  of the astrospheres to assist in inferring mass loss rates from the
  absorption.  The analyzed stars are $\epsilon$~Eri
  ($\dot{M}=30~\dot{M}_{\odot}$), 61~Cyg~A ($\dot{M}=0.5~\dot{M}_{\odot}$),
  36~Oph~AB ($\dot{M}=15~\dot{M}_{\odot}$), and 40~Eri~A
  ($\dot{M}<5~\dot{M}_{\odot}$).
\item[2.] The astrospheres vary greatly in size depending on the stellar
  mass loss rate and the interstellar wind velocity seen by the star.
  The astrosphere of 61~Cyg~A has an upwind bow shock distance of only
  30~AU.  In contrast, the upwind bow shock distance for $\epsilon$~Eri is
  about 1600~AU, with an apparent angular width of about
  $42^{\prime}$ as seen from Earth, making $\epsilon$~Eri's
  astrosphere larger than the size of the full Moon.
\item[3.] Combining our mass loss measurements with previous measurements,
  we study how the mass loss rates of cool main sequence stars depend on
  activity.  The solar-like GK dwarfs show a correlation between mass loss
  rate (per unit surface area) and X-ray surface flux that can be described
  as a power law: $\dot{M}\propto F_{x}^{1.15\pm 0.20}$.
\item[4.] Both Proxima~Cen (M5.5~Ve) and the RS~CVn system $\lambda$~And
  (G8~IV-III+M~V) have mass loss rates significantly lower than would be
  predicted by the mass-loss/activity relation defined by the solar-like
  GK dwarfs, although we note that the astrospheric detection for
  $\lambda$~And must be considered uncertain.
\item[5.] We combine our power law relation for mass loss and activity
  with rotation/activity and age/rotation relations from \citet{tra97} to
  derive a mass-loss/age relation for solar-like stars:
  $\dot{M}\propto t^{-2.00\pm 0.52}$.  This relation suggests that the
  solar wind may have been as much as 1000 times stronger in the distant
  past.
\item[6.] Our results are consistent with theoretical descriptions
  of the magnetic braking process that slows stellar rotation only if
  disk-averaged stellar magnetic fields decline at least inversely with
  time (i.e., $B\propto t^{\alpha}$, where $\alpha<-1$).
\item[7.] Our suggestion that the young solar wind may have been
  significantly more massive than it is now could have important
  ramifications for the history of planetary atmospheres in our solar
  system.  The more massive young solar wind suggested by our analysis is
  still not massive enough to resolve the so-called ``faint young Sun''
  paradox, but the stronger wind does make it more likely that erosion
  by the solar wind has dramatically changed the properties of
  planetary atmospheres, with the Martian atmosphere being a particularly
  interesting case.
\end{description}

\acknowledgments

We would like to thank M.\ Giampapa for useful comments on the manuscript.
Support for this work was provided by NASA through grants NAG5-9041 and
S-56500-D to the University of Colorado, and through grant number
GO-08237.01-A from the Space Telescope Science Institute, which is operated
by AURA, Inc., under NASA contract NAS5-26555.  G.~P.~Z.\ and H.~-R.~M.\
would also like to acknowledge support from NSF-DOE grant ATM-0296114.

\clearpage

\begin{deluxetable}{lccccccccc}
\tabletypesize{\small}
\tablecaption{Mass Loss Measurements}
\tablecolumns{9}
\tablewidth{0pt}
\tablehead{
  \colhead{Star} & \colhead{Spectral} & \colhead{$d$} &
    \colhead{Surf.\ Area} & \colhead{Log L$_{x}$} & \colhead{$V_{ISM}$} &
    \colhead{$\theta$} & \colhead{$\log {\rm N(H~I)}$} &
    \colhead{$\dot{M}$} & \colhead{Refs.} \\
  \colhead{} & \colhead{Type} & \colhead{(pc)} & \colhead{(A$_{\odot}$)} &
    \colhead{} & \colhead{(km~s$^{-1}$)} & \colhead{(deg)} & \colhead{} &
    \colhead{($\dot{M}_{\odot}$)} & \colhead{}}
\startdata
$\alpha$ Cen  & G2 V+K0 V   & 1.3 & 2.37 & 27.34 & 25 & 79 &15.24& 2& 1,2,3\\
Prox Cen      & M5.5 V      & 1.3 & 0.026& 27.23 & 25 & 79 &...  &$<0.2$& 3\\
$\epsilon$ Eri& K1 V        & 3.2 & 0.62 & 28.32 & 27 & 76 &15.82& 30& 4,10\\
61 Cyg A      & K5 V        & 3.5 & 0.45 & 27.26 & 86 & 46 &14.11&0.5& 5,10\\
$\epsilon$ Ind& K5 V        & 3.6 & 0.50 & 27.18 & 68 & 64&14.25&0.5& 6,7,8\\
40 Eri A\tablenotemark{a}&K1 V& 5.0&0.64 & 27.61 &127 & 59 &... &$<5$& 5,10\\
36 Oph        & K1 V+K1 V   & 5.5 & 0.88 & 28.28 & 40 &134 &16.10& 15& 9,10\\
$\lambda$ And\tablenotemark{a}&G8 IV-III+M V&26&55&30.53&53&89&15.20&5& 6,7,8\\
\enddata
\tablenotetext{a}{Uncertain detection.}
\tablerefs{(1) Linsky \& Wood 1996. (2) Gayley et al.\ 1997. (3) Wood
  et al.\ 2001. (4) Dring et al.\ 1997. (5) Wood \& Linsky 1998. (6) Wood
  et al.\ 1996. (7) M\"{u}ller et al.\ 2001a. (8) M\"{u}ller et al.\ 2001b.
  (9) Wood et al.\ 2000a. (10) This paper.}
\end{deluxetable}

\clearpage

\begin{figure}
\plotfiddle{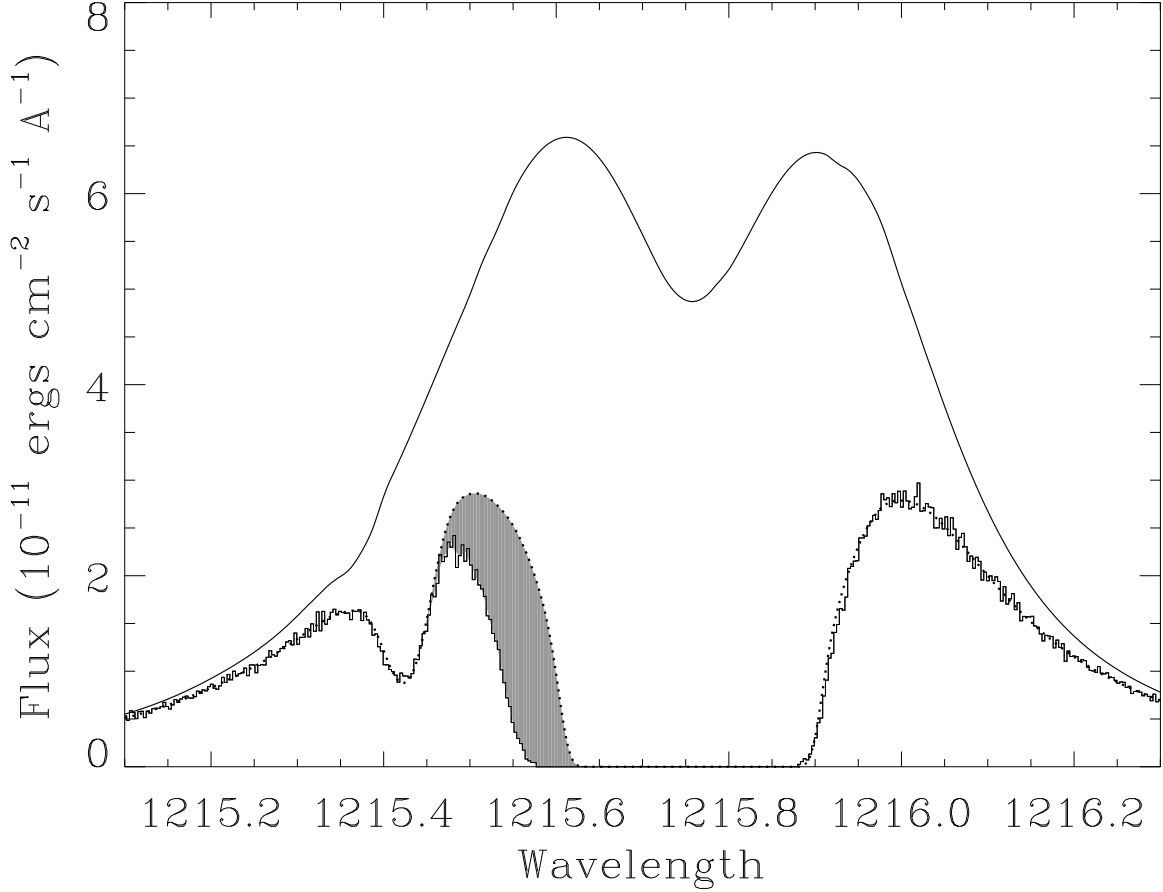}{3.5in}{90}{75}{75}{280}{0}
\caption{HST/GHRS Ly$\alpha$ spectrum of $\epsilon$~Eri, showing broad
  H~I absorption at 1215.7~\AA\ and narrow D~I absorption at 1215.4~\AA.
  The upper solid line is the assumed intrinsic stellar emission line, and
  the dotted line is the profile after ISM absorption alone, derived by
  forcing consistency between the ISM H~I and D~I absorption.  The excess
  H~I absorption on the blue side of the line (shaded region) is
  astrospheric absorption.}
\end{figure}

\clearpage

\begin{figure}
\plotfiddle{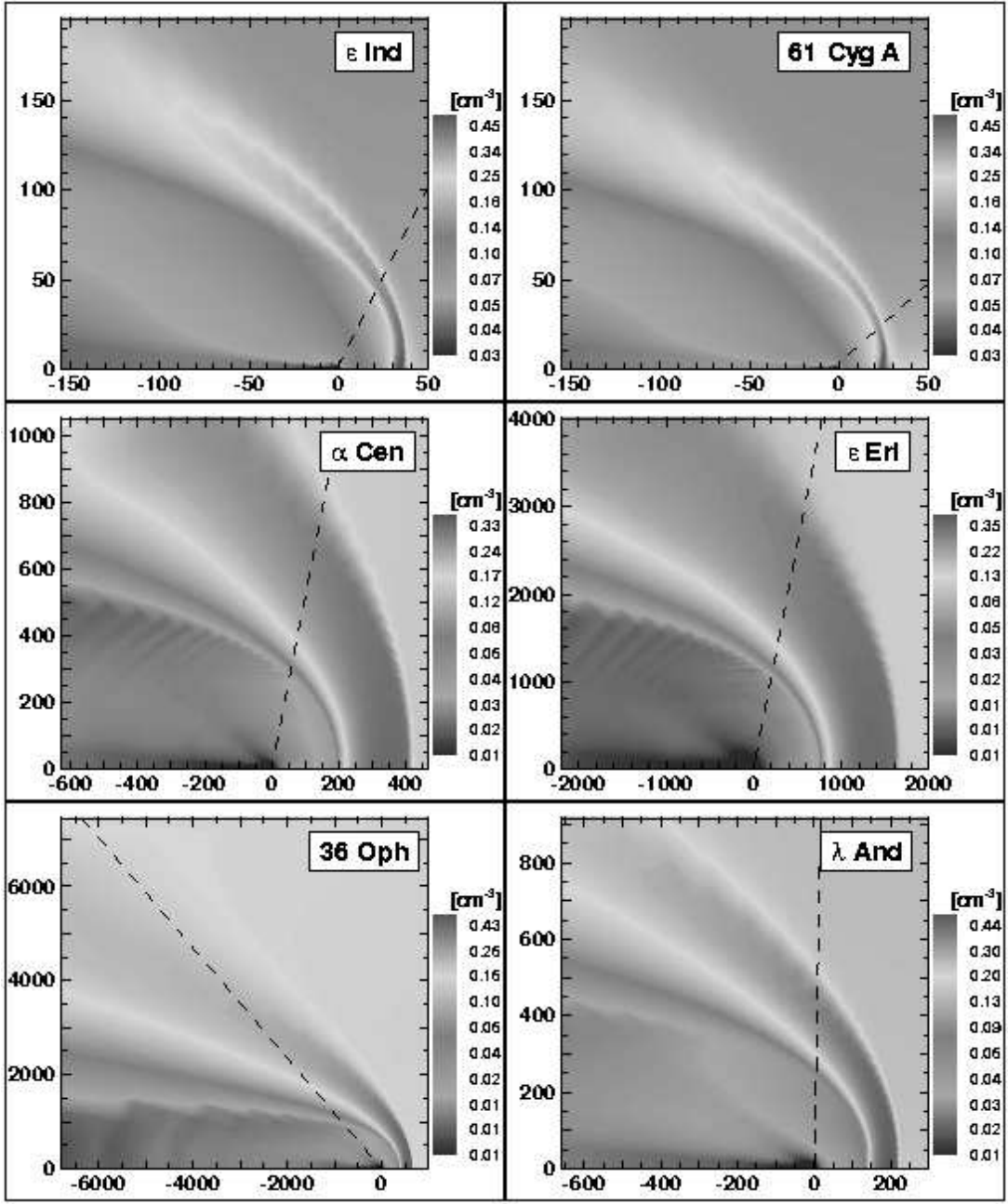}{6.5in}{0}{100}{100}{-300}{-150}
\caption{Maps of H~I density from hydrodynamic models of stellar
  astrospheres.  The models shown are the ones that lead to the best fits
  to the data in Fig.~3.  The distance scale is in AU.  The star is at
  coordinate (0,0) and the ISM wind is from the right.  The dashed lines
  indicate the Sun--star line of sight.}
\end{figure}

\clearpage

\begin{figure}
\plotfiddle{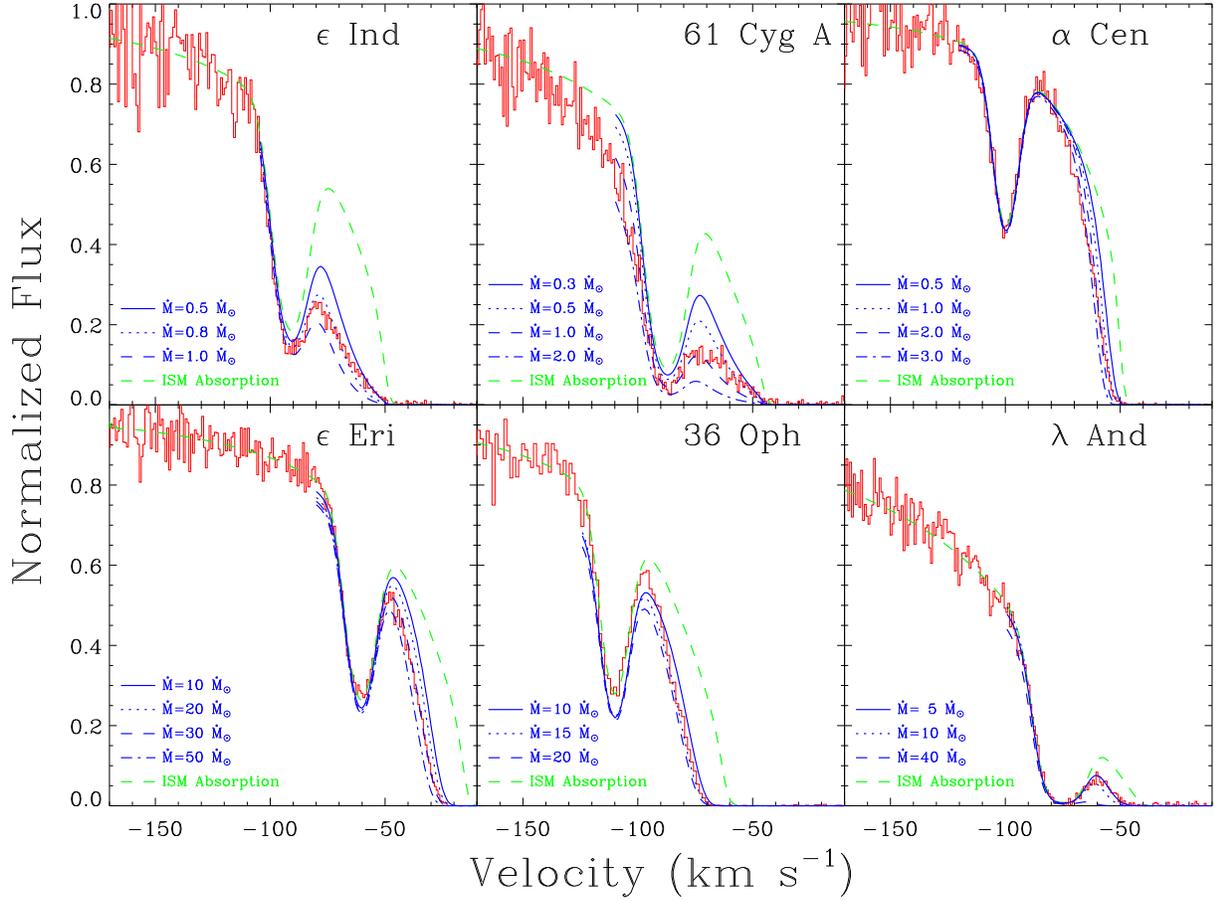}{3.5in}{90}{75}{75}{280}{0}
\caption{Closeups of the blue side of the H~I Ly$\alpha$ absorption lines
  for all stars with detected astrospheric absorption, plotted on a
  heliocentric velocity scale.  Narrow D~I ISM absorption is visible in
  all the spectra just blueward of the saturated H~I absorption.  Green
  dashed lines indicate the interstellar absorption alone, and blue lines
  in each panel show the additional astrospheric absorption predicted by
  hydrodynamic models of the astrospheres assuming various mass loss rates.}
\end{figure}

\clearpage

\begin{figure}
\plotfiddle{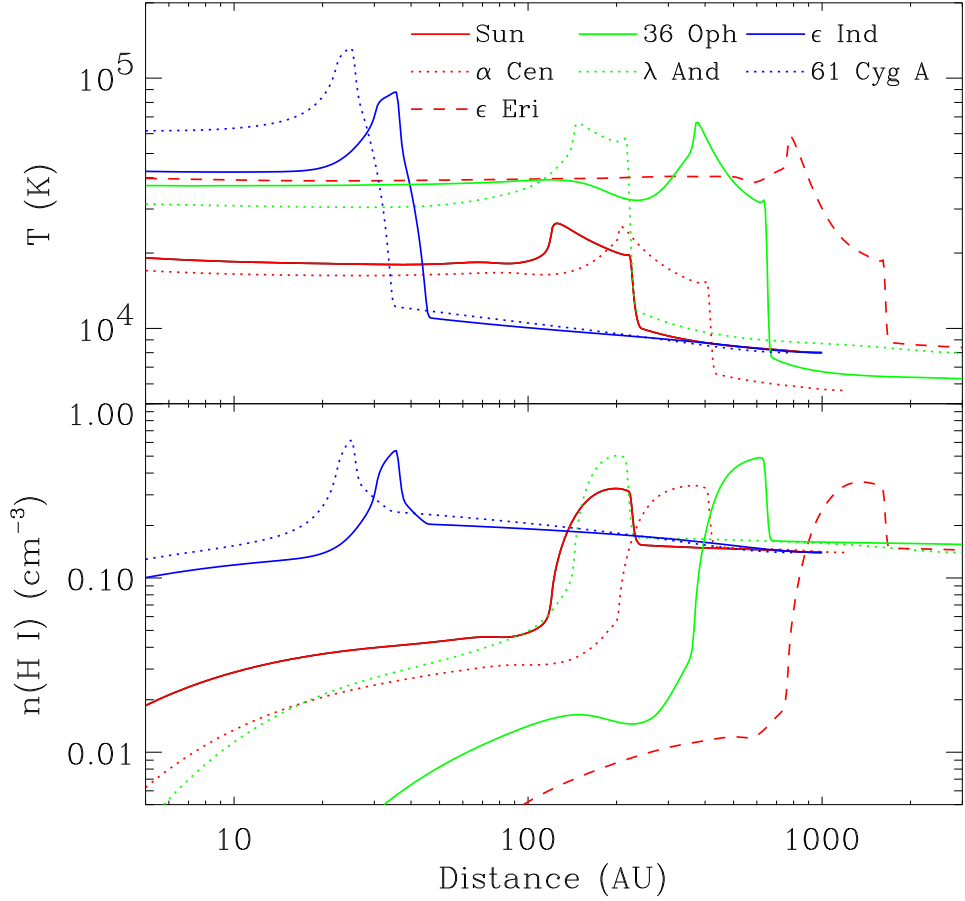}{3.5in}{90}{75}{75}{300}{-20}
\caption{Variation of the H~I temperature and density with distance from
  the star in the upwind direction ($\theta=0^{\circ}$) based on the
  ``best-fit'' models in Fig.~2, and on the best solar model from Wood
  et al.\ (2000b).}
\end{figure}

\clearpage

\begin{figure}
\plotfiddle{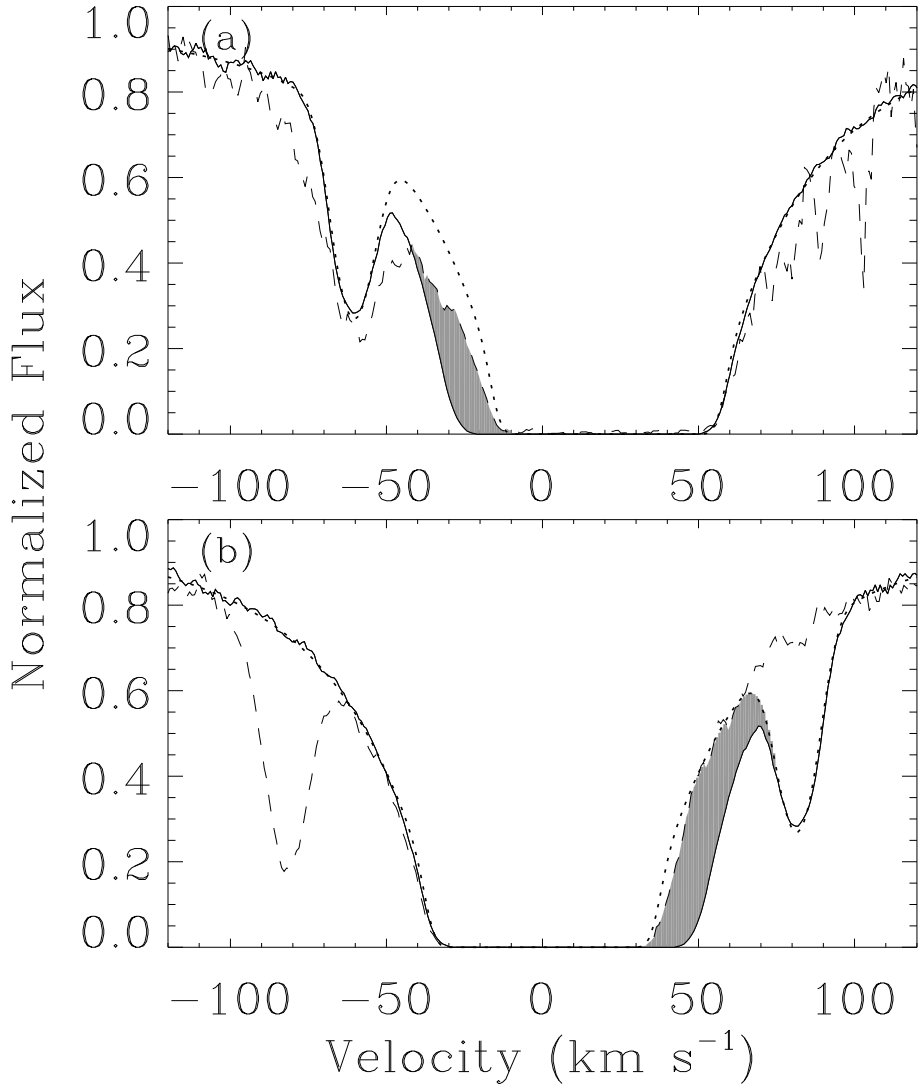}{5.5in}{0}{80}{80}{-270}{0}
\caption{(a) The observed H~I Ly$\alpha$ line of $\epsilon$~Eri (solid
  line) and the Ly$\alpha$ profile with only ISM absorption (dotted line)
  for $\epsilon$~Eri from Fig.~1 are compared with
  the H~I absorption observed toward 40~Eri (dashed line), which samples
  a heliospheric angle similar to $\epsilon$~Eri.  The shaded region is the
  excess H~I absorption seen toward $\epsilon$~Eri that is not seen toward
  40~Eri, demonstrating that the excess absorption cannot be heliospheric,
  supporting an astrospheric interpretation.  The velocity scale is
  heliocentric.  (b) The $\epsilon$~Eri profiles from (a) are compared with
  the H~I Ly$\alpha$ line absorption observed toward 31~Com (dashed line),
  which samples a heliospheric angle similar to the astrospheric angle
  sampled toward $\epsilon$~Eri.  The profiles are plotted on a velocity
  scale centered on the ISM rest frame and the $\epsilon$~Eri profile is
  reversed about this rest frame in order to compare the astrospheric
  absorption seen toward $\epsilon$~Eri with the upper limit for the amount
  of heliospheric absorption provided by the 31~Com data.  There is clearly
  more astrospheric absorption for $\epsilon$~Eri than heliospheric
  absorption toward 31~Com (shaded region), implying that $\epsilon$~Eri
  must have a larger mass loss rate than the Sun.
  The data in both panels have been smoothed for clarity.}
\end{figure}

\clearpage

\begin{figure}
\plotfiddle{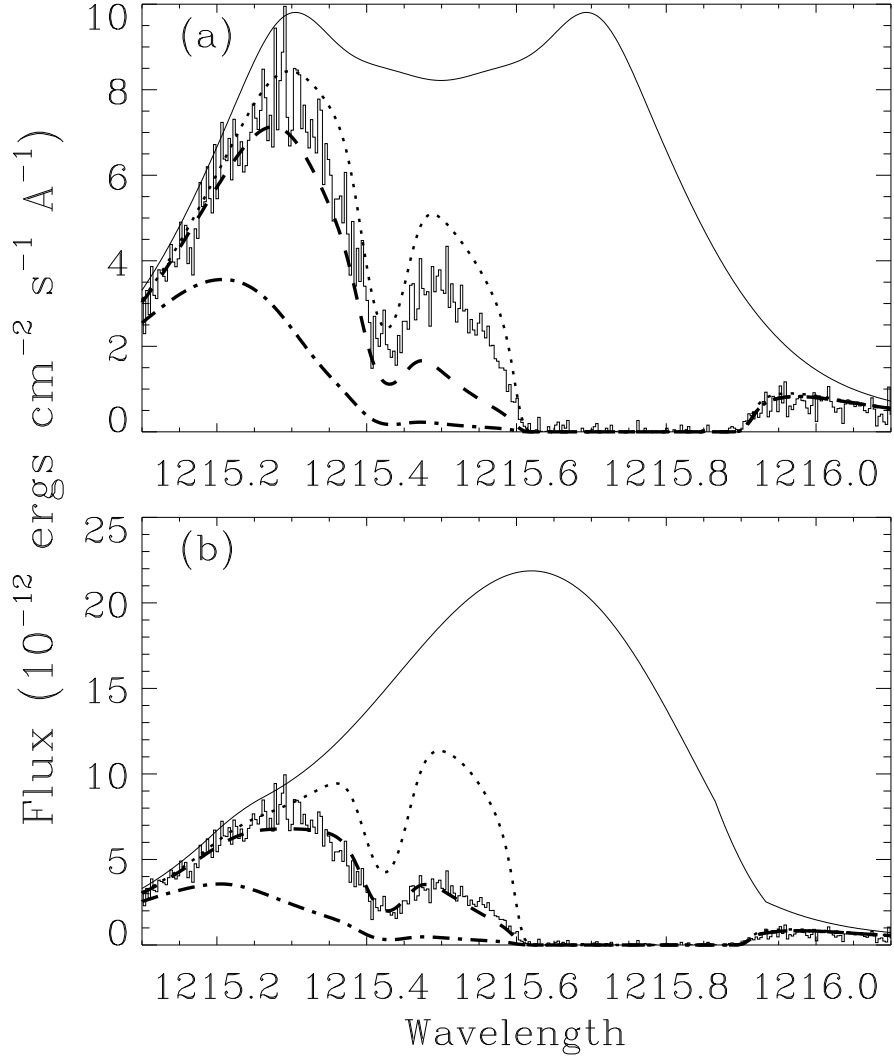}{5.0in}{0}{80}{80}{-270}{0}
\caption{(a) HST Ly$\alpha$ spectrum of 40~Eri~A, showing broad
  H~I absorption at 1215.8~\AA\ and narrow D~I absorption at 1215.4~\AA.
  The upper solid line is the assumed intrinsic stellar emission line, and
  the dotted line shows the Ly$\alpha$ profile after only ISM absorption.
  The excess H~I absorption on the blue side of the line is astrospheric
  absorption.  The dashed and dot-dashed lines are predicted astrospheric
  absorption from astrospheric models with mass loss rates of
  $\dot{M} = 1~\dot{M}_{\odot}$ and $\dot{M} = 5~\dot{M}_{\odot}$,
  respectively. (b) Similar to (a), but with a different assumed stellar
  Ly$\alpha$ profile.  The $\dot{M} = 1~\dot{M}_{\odot}$ model (dashed
  line) now fits the data much better.}
\end{figure}

\clearpage

\begin{figure}
\plotfiddle{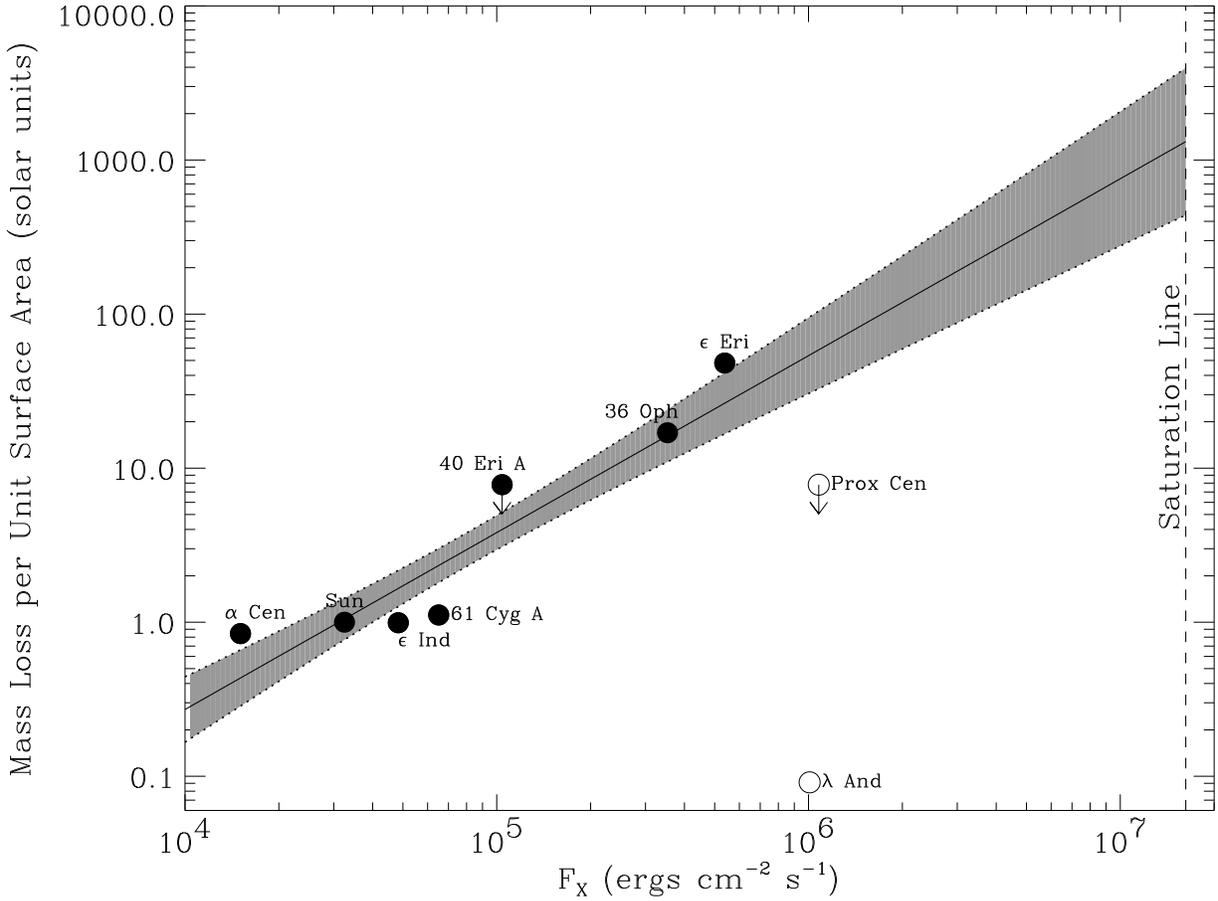}{3.5in}{90}{70}{70}{285}{0}
\caption{Measured mass loss rates (per unit surface area) plotted
  versus X-ray surface flux.  A power law has been fitted to the solar-like
  GK dwarfs (filled circles), and the shaded region is the estimated
  uncertainty in the fit.  Proxima Cen (M5.5 Ve) and $\lambda$~And
  (G8 IV-III+M V) appear to be inconsistent with this relation.  The
  saturation line represents the maximum F$_{X}$ value observed from
  solar-like stars.}
\end{figure}

\clearpage

\begin{figure}
\plotfiddle{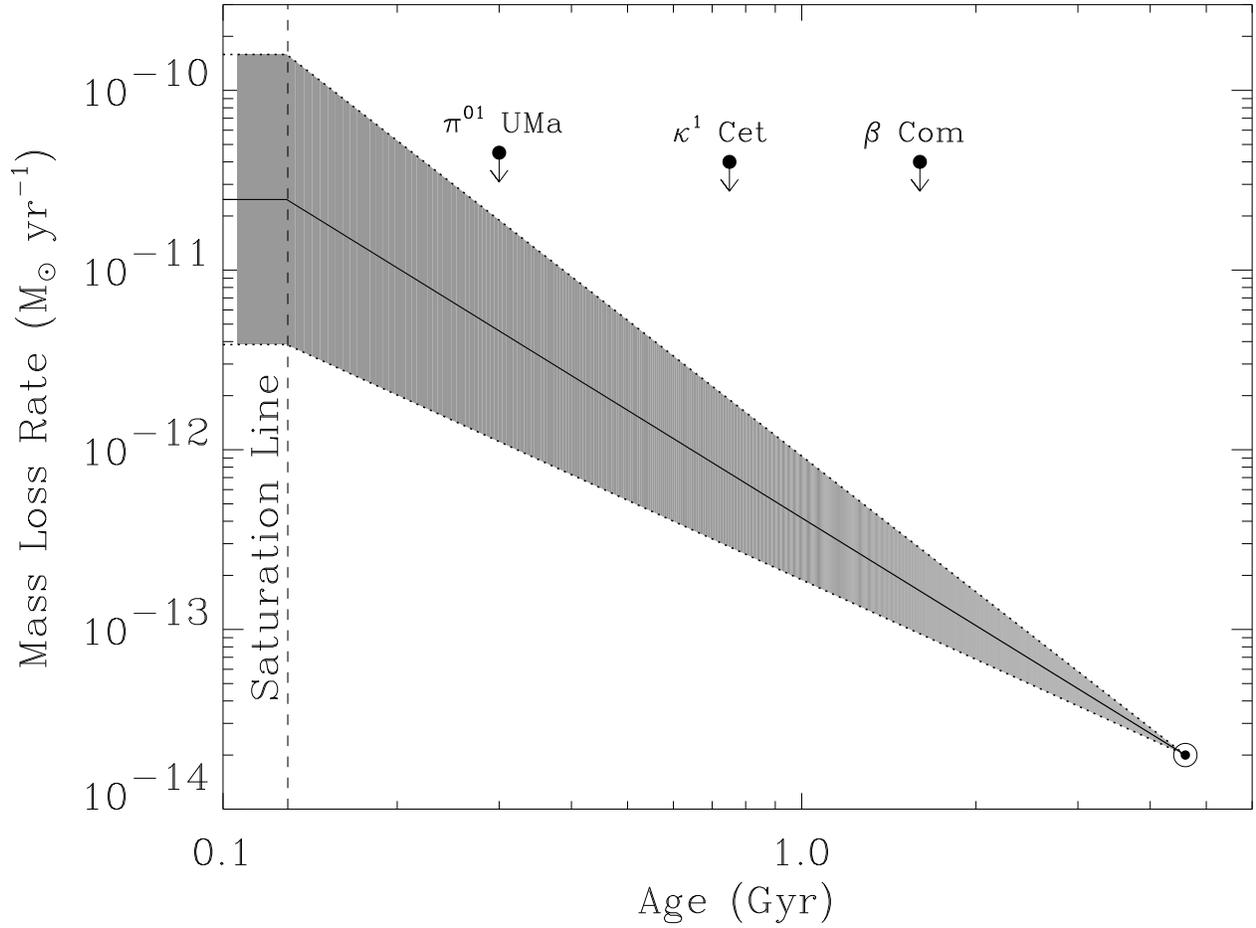}{3.5in}{90}{70}{70}{285}{0}
\caption{The mass loss history of the Sun suggested by the power law
  relation from Fig.~7.  The upper limits are based on radio nondetections
  of three solar-like stars (Gaidos et al.\ 2000).}
\end{figure}

\clearpage

\begin{figure}
\plotfiddle{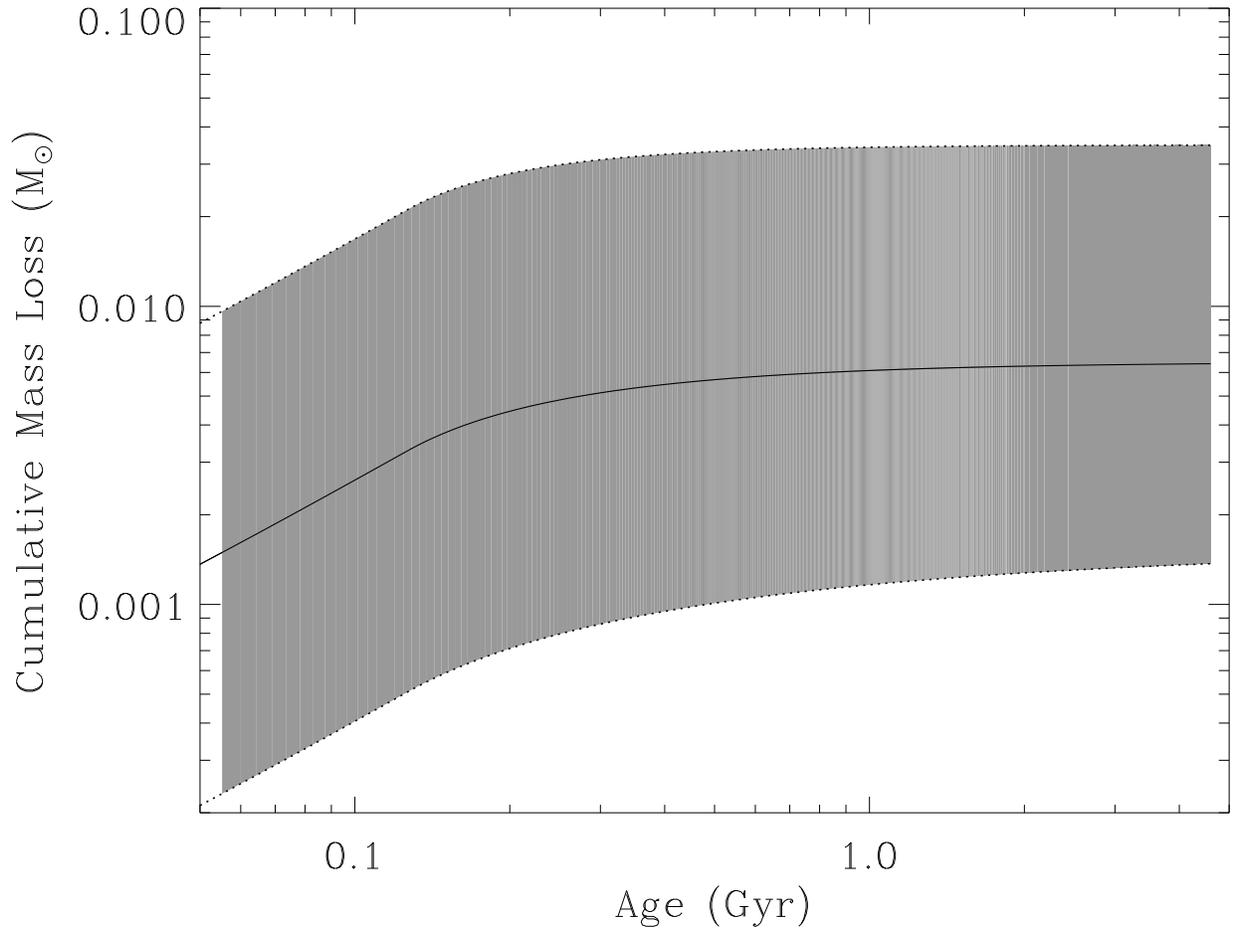}{3.5in}{90}{70}{70}{285}{0}
\caption{Cumulative mass loss for the Sun as a function of time, based
  on the mass loss history of the solar wind in Fig.~8.}
\end{figure}

\end{document}